\documentclass[apj]{emulateapj}
\usepackage{apjfonts}
\usepackage{amsmath}
\usepackage{natbib}

\submitted{Accepted for publication in the Astrophysical Journal}

\usepackage{graphicx}
\usepackage[figuresright]{rotating} 
\usepackage{float}       
\usepackage{afterpage}   

\newcommand {\BR}{bremsstrahlung}
\newcommand {\SY}{synchrotron}
\newcommand {\SO}{SN~1006}
\newcommand {\asca}{\emph{ASCA}}
\newcommand{\integral}{\emph{INTEGRAL}}
\newcommand{\wsim}{\ensuremath{\sim}}
\newcommand{\rxte}{\emph{RXTE}}
\newcommand{\xmm}{\emph{XMM-Newton}}
\newcommand{\chandra}{\emph{Chandra X-ray Observatory}}

\begin{document}


\title{X-ray observations of SN 1006 with \emph{INTEGRAL}}

\author{E. Kalemci\altaffilmark{1,2},
        S. P. Reynolds\altaffilmark{3},
        S. E. Boggs\altaffilmark{1,4},
        N. Lund\altaffilmark{5},
        J. Chenevez\altaffilmark{5},
    M. Renaud\altaffilmark{6,7},
    J. Rho\altaffilmark{8}
}

\altaffiltext{1}{Space Sciences Laboratory, 7 Gauss Way, University of
California, Berkeley, CA, 94720-7450, USA.}

\altaffiltext{2}{Sabanc\i\ University, Orhanl\i -Tuzla 34956,
\.Istanbul, Turkey.}

\altaffiltext{3}{Department of Physics, NC State University, 2700
Stinson Drive, Box 8202, Raleigh, NC 27695, USA.}

\altaffiltext{4}{Department of Physics, University of California,
366 Le Conte Hall, Berkeley, CA, 94720-7300, USA.}

\altaffiltext{5}{Danish National Space Center, Juliane Maries Vej
30, DK-2100 Copenhagen $\emptyset$, Denmark.}

\altaffiltext{6}{Service d'Astrophysique, CEA-Saclay, 91191,
Gif-Sur-Yvette, France}

\altaffiltext{7}{APC-UMR 7164, 11 place M. Berthelot, 75231 Paris,
France}

\altaffiltext{8}{SIRTF Science Center, California Institute of
Technology Mail Stop 220-6, Pasadena, CA 91125}


\begin{abstract}

The remnant of the supernova of 1006 AD, the remnant first showing
evidence for the presence of X-ray synchrotron emission from
shock-accelerated electrons, was observed for \wsim 1000 ksec with
\integral\ for the study of electron acceleration to very high
energies. The aim of the observation was to characterize the
synchrotron emission, and attempt to detect non-thermal bremsstrahlung,
using the combination of IBIS and JEM-X spatial and spectral
coverage. The source was detected with JEM-X between 2.4 and 8.4 keV bands, and
 not detected with either ISGRI or SPI above 20 keV. The ISGRI upper limit is
about a factor of four above current model predictions, but
confirms the presence of steepening in the power-law extrapolated
from lower energies ($<4$ keV).

\end{abstract}




\section{Introduction}\label{sec:intro}

Supernova remnants (SNRs) have long been thought to be the primary
site of Galactic cosmic ray acceleration up to the ``knee''
feature in the integrated cosmic-ray spectrum near 3000 TeV, as
the supernova shocks are one of the few mechanisms that could
provide enough energy to support this population \citep{Dyer01}.
However, many features of the acceleration process, including
injection physics, efficiency, and maximum electron and ion
energies, are not yet clear.  Hard X-ray observations (above 10
keV) of SNRs may cast light on this poorly understood process.

High energy electrons in SNRs produce X-rays via two mechanisms,
non-thermal \\
\BR\ and \SY \ radiation. Electrons can also produce
gamma-rays up to TeV energies by inverse Compton (IC) scattering of
any photons present, such as the cosmic microwave-background (CMB).
In addition, relativistic protons can produce gamma-rays from the
decay of $\pi^{0}$ particles from inelastic ion-ion collision. All
these processes have been extensively modeled for SNRs by various
groups \citep{Sturner97,Gaisser98, Baring99}. In hard X-rays, \SY\
radiation from the tail of the electron distribution may compete
with non-thermal \BR\ from the very lowest-energy accelerated
electrons.

\SO\ has been the prototype laboratory for the study of electron
acceleration to high energies in shocks.  X-rays from this object were
first reported by \cite{Winkler76}. The earlier featureless spectrum
\citep{Becker80} was modeled as the loss-steepened extrapolation of
the radio \SY\ spectrum by \cite{Reynolds81}. Later, observations by
\emph{ASCA} \citep{Koyama95} showed that the limbs have featureless
spectra well described by power-laws, whereas the interior has a
thermal, line-dominated spectrum. The source was also observed with
\rxte, and \cite{Dyer01} showed that elaborate synchrotron emission
models \citep{Reynolds96, Reynolds98} fit the combined \rxte\ -
\emph{ASCA} spectrum reasonably well.

Electrons producing keV \SY\ emission could also produce very
high-energy photons (in the TeV range) by IC upscattering of CMB
photons \citep{Pohl96}. The TeV flux depends on the electron
distribution, and in conjunction with the synchrotron flux, a mean
magnetic field strength of the remnant can be deduced. A detection
of the northeast (NE) limb of \SO\ was reported in ground-based TeV
observations by \emph{CANGAROO-I} \citep{Tanimori98}. It is
interesting that only one limb was detected, even though the X-ray
spectra of the two limbs are similar \citep{Allen01, Dyer04}. To
have a discrepancy in the TeV band, the electron spectra, magnetic
field strengths, or \SY\ and IC emission-angle distributions of the
two rims would have to be different \citep{Allen01}, but in a way
that does not produce significant differences in the X-ray band.

The nature of X-ray emission from \SO\ above 10 keV is still
 uncertain. Below 10 keV, \SY\ emission is the most plausible
explanation. For \SY\ radiation, the quantitative inferences apply
only to the exponential cutoff of the electron distribution. On the
other hand, most of the accelerated electrons and much of their total
energy reside in the lowest energy non-thermal electrons, whose \BR\
emission could become dominant above 30 keV. In principle, the
\integral\ observatory can examine the effects of both the lowest and
highest energy non-thermal electrons by distinguishing the \SY\ and
\BR\ emission with its imaging and spectral capabilities.

\cite{Reynolds99} modeled bremsstrahlung and synchrotron emission
from \SO . Synchrotron hard X-rays should be concentrated in two
bright opposing limbs like the radio emission, and should dominate
the emission below 30 keV. The images and spectra taken with
\asca\ \citep{Koyama95}, \chandra\ \citep{Long03}, and \xmm\
\citep{Rothenflug04} confirm this below 10 keV. The non-thermal
bremsstrahlung, resulting from slightly supra-thermal
shock-accelerated electrons interacting with thermal ions, is
likely to be more symmetrically distributed, and will dominate at
some energy between 30 and 300 keV. The bremsstrahlung flux will
scale with the product of the thermal gas density $n_{\rm th}$ and
the relativistic-electron density $n_{\rm e,rel}$.   The former
can be constrained by observations of thermal X-ray emission,
while the latter can be deduced from radio synchrotron fluxes if
the magnetic field is known.

The TeV spectrum reported from \emph{CANGAROO-I} \citep{Tanimori98}
could be well-described by IC upscattered CMB photons, using a
power-law electron spectrum with an exponential cutoff as described
in \cite{Dyer01}.  This fit gave a post-shock magnetic field of
about 10 $\mu$G. However, this result is now called into question by
the observations of \emph{H.E.S.S.}, which did not detect the source
despite better sensitivity compared to \emph{CANGAROO-I}. The TeV
upper limits from \emph{H.E.S.S.} are about a factor of 10 below the
\emph{CANGAROO-I} results\footnote{Recent measurements by
\emph{CANGAROO-III} has also claimed null result on \SO\
\citep{Tanimori05ICRS}.} \citep{Aharonian05}. These limits constrain
IC upscattering of CMB photons by the same electrons that produce
X-ray synchrotron emission: tighter limits mean fewer electrons, a
higher magnetic field (Aharonian et al.~place a lower limit of 25
$\mu G$ on the post-shock magnetic field), and less non-thermal \BR.
Lowering the possible IC-CMB flux by a factor of 10 directly lowers
the allowable relativistic-electron density by the same factor, and
hence lowers the predicted bremsstrahlung flux by an order of
magnitude.

Even if bremsstrahlung is not detected, the detailed shape of the
steepening synchrotron spectrum can provide information on the
physical process causing the cutoff in the electron spectrum. This
is crucial information for understanding the acceleration of
cosmic rays, since if the cutoff is due to radiative losses on
electrons, the proton spectrum might extend to much higher
energies, perhaps as high as the knee.  However, if the finite
remnant age (or size) or some change in diffusive properties of
the upstream medium causes the cutoff, it should also cut the
proton spectrum off at a similar energy (between 10 and 100 TeV;
\citealt{Dyer01}), far below the ``knee'' energy.  Detailed models
show subtle but potentially distinguishable differences in the
shape of the synchrotron spectrum above 10 keV, with loss-limited
spectra (due to higher magnetic fields) being somewhat harder.

Our group has observed \SO\ for \wsim 1000 ks with the \integral\
Observatory \citep{WinklerC03} in AO-1 with the main aim of
detecting and characterizing synchrotron emission, and
distinguishing synchrotron and non-thermal bremsstrahlung emission
by comparing the IBIS/ISGRI and JEM-X images to the model images. In
this work, we will discuss the results of the analysis of the
\integral\ data, and place limits on the \SY\ and \BR\ emission from
\SO.


\section{Observations and Analysis}\label{sec:obs}

The \integral\ observations took place in two sets. The \wsim 250
ks first set (``Set I'') was conducted early in the mission, between Jan,
11, 2003, and Jan, 20, 2003, corresponding to \integral\ revolutions 30 and 32.
 The \wsim 750 ks second set (``Set II'') was conducted between Jan, 20, 2004
and Jan, 30, 2004, during revolutions 155-158. These two sets have
different observational characteristics for different instruments as
explained below, and only the data from Set II are used for this
paper. We did not use SPI \citep[Spectrometer on \integral,
][]{Vedrenne03}, as ISGRI (see \S~\ref{subsec:isgrian} for more
information) places much stricter limits in the hard X-ray band.
Before the general analysis for all instruments, we filtered out the
pointings with high Anti-Coincidence Shield rates, mostly occurring
during the entry and exit of the radiation belts.

\subsection{The JEM-X analysis}\label{subsec:jemxan}

The Joint European X-ray Monitor, JEM-X, consists of two identical
high pressure imaging micro-strip gas chambers, and makes
observations simultaneously with the main instruments on
\integral, albeit with a narrower fully-coded field of view of
4.8$^{\circ}$. The energy band is 3--35 keV and the angular
resolution is 3.35$^{\prime}$ \citep{Lund03}. Due to a problem
with eroding anodes, the high voltage in the JEM-X detectors was
lowered, and a new background rejection criterion was implemented
after the launch. One of the detector pair is being kept in a safe
state, and during our observations only JEM-X 2 was operational.

Set I was conducted before the new background rejection criteria
were implemented in JEM-X, and therefore was not included in this
analysis. We note that, although the total exposure for Set II is
\wsim 750 ks, the effective exposure time of the central object is
approximately 250 ks due to the vignetting of the JEM-X instrument
during the 25 point dither. We have obtained JEM-X images in 4
energy bands using the \emph{JEM-X Midisky offline software package}
available from the DNSC \citep{Lund04}. These energy
bands\footnote{The low energy threshold is variable over different
parts of the JEM-X detector.} are 2.4--4.2 keV, 4.2--8.4 keV,
8.4--14 keV, and 14--35 keV. The images from each pointing are then
mosaicked using the \emph{mosaic-weight} program \citep{Chenevez04}.

\begin{figure}
\plotone{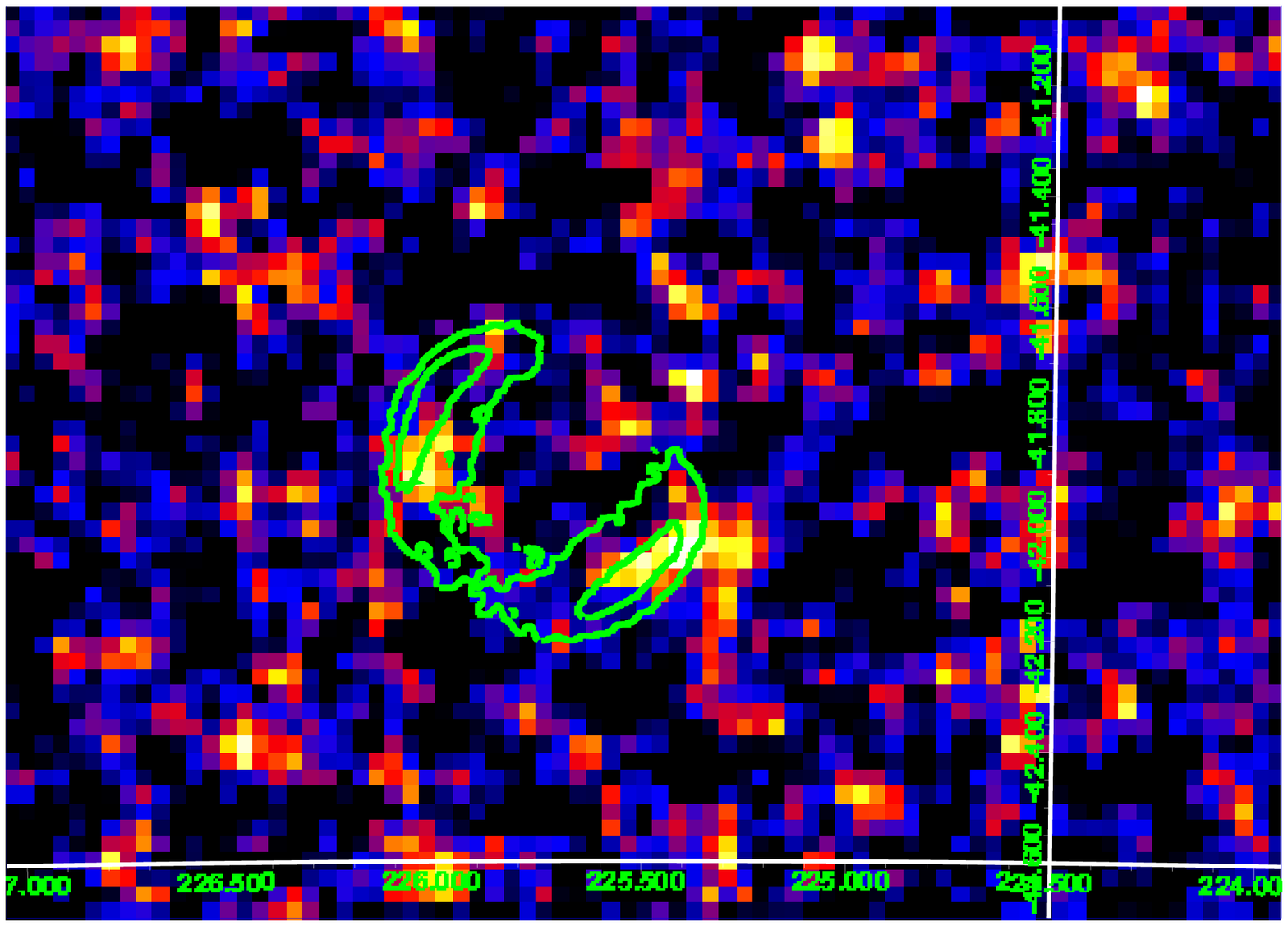} \plotone{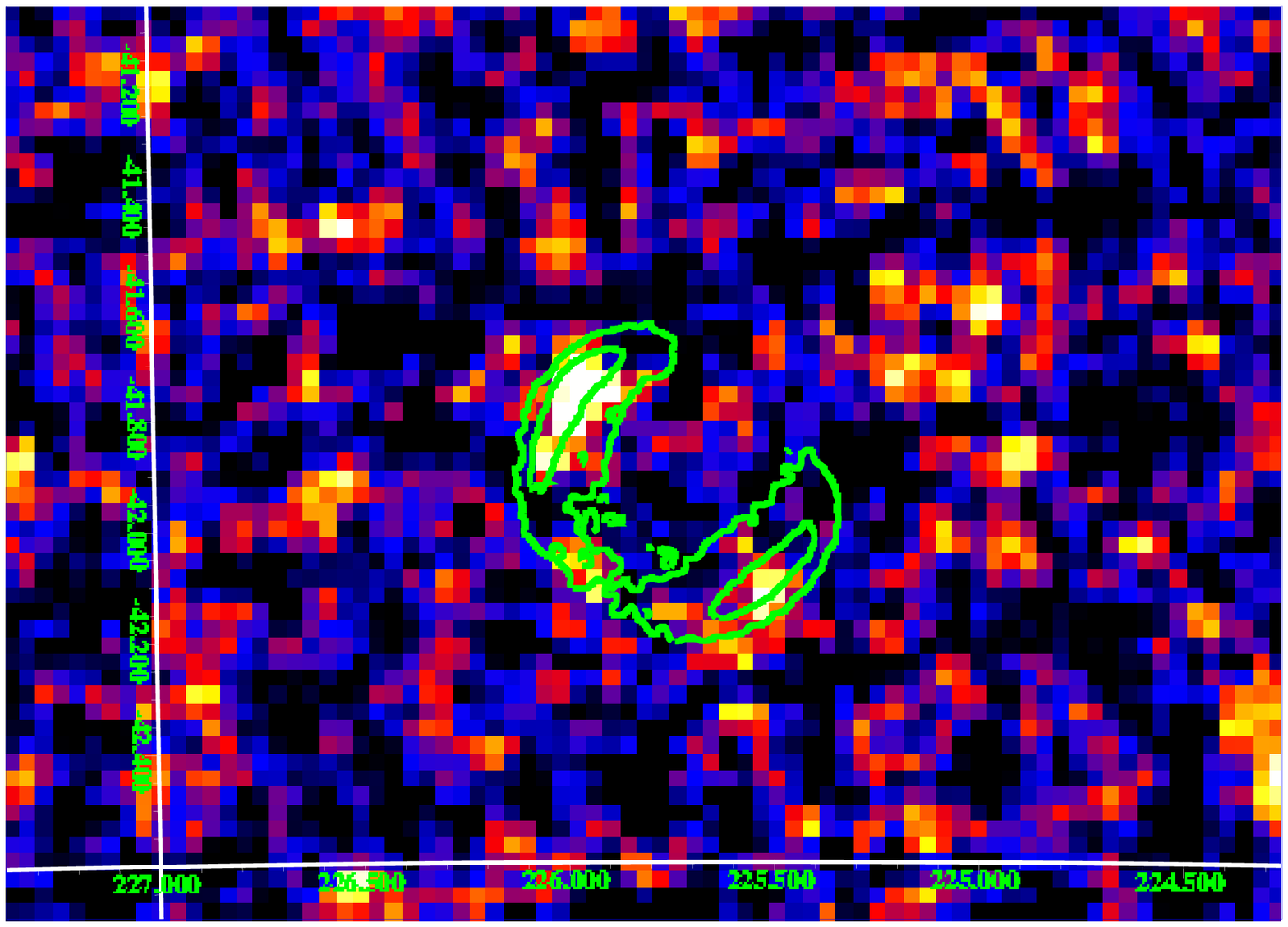} \plotone{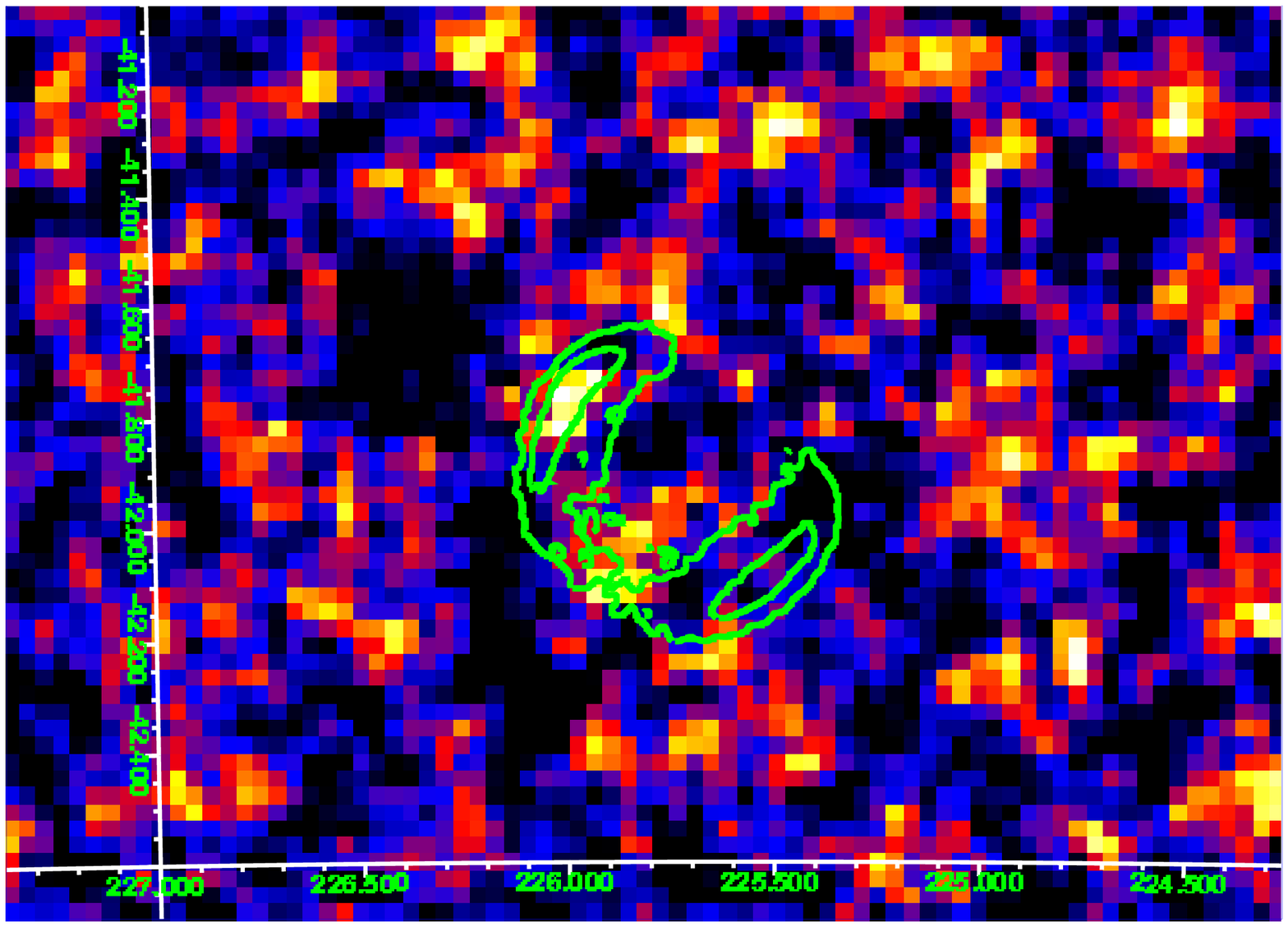}
\caption{\label{fig:jemim} JEM-X reconstructed images of \SO\ in
three energy bands. Top: 2.4--4.2 keV, middle: 4.2--8.4 keV band,
bottom: 8.4-14 keV. The \asca\ contours are overlaid. The detection
significances, fluxes, and upper limit fluxes are shown in
Table~\ref{table:jemxsum}. (The electronic version has this figure
in color.) }
\end{figure}

The fluxes and the significance values shown in
Table~\ref{table:jemxsum} are derived using the inner ASCA
contours enclosing the NE and SW limbs to define two shape
templates. The NE template is constructed with 23 pixels inside
the inner ASCA contour seen in Fig.~\ref{fig:jemim} on the left.
The SW template is also constructed similarly, using the 19 pixels
inside the inner ASCA contour on the right. The count excesses in
the JEM-X mosaic images are then determined inside these two
regions. The relevant noise figures are derived by defining a
number of non-overlapping regions with the reference templates
within a 60 $\times$ 60 pixel field (90 $\times$ 90 arcminutes)
centered on SN1006. 114 NE-templates, each containing 23 pixels,
can be arranged within the 60 $\times$ 60 field. For the
SW-template the corresponding numbers are 112 regions, each with
19 pixels. The statistical properties of the excess counts in
these two sets of regions are used to derive RMS-noise of the
background for regions of the two shapes. The signal-to-noise
ratio derived in this way will obey normal statistics because the
region templates are defined independently of the JEM-X data.

The fluxes are derived by comparing the excesses in the \SO\ mosaics
with corresponding excesses in mosaic images of the Crab Nebula
obtained with the same INTEGRAL dither pattern, and correcting for
the difference in the effective observation time.

\subsection{The ISGRI analysis}\label{subsec:isgrian}

One of the two main instruments on \integral, IBIS (Imager on Board
the INTEGRAL Satellite, \citealt{Ubertini03}), consists of two
cameras. The \integral\ Soft Gamma-ray Imager, ISGRI, is the
low-energy camera of the IBIS telescope \citep{Lebrun03}. It has a
large sensitive area of 2621 cm$^2$ made up of 16384 CdTe pixels.
The angular resolution is \wsim 13$^{\prime}$ \citep{Gros03}, and
the fully-coded field of view is 9$^{\circ}$. The energy range is 20
keV -- 10 MeV. The older background maps for Set I observations
resulted in much noisier images compared to Set II images, and
therefore we limit the analysis to the data from Set II. We used OSA
4.2 \citep{Goldwurm03} standard programs to obtain images in 20--40
keV and 40--100 keV bands.

Since the ISGRI system point spread function is \wsim 13$^{\prime}$,
\SO\ (\wsim 30$^{\prime}$ diameter)
appears as an extended object to the imager.
Estimation of the flux of such sources with a coded-mask instrument is
complicated, since the mask patterns used in gamma-ray astronomy are optimized
for point sources. One needs to use simulations to obtain the effect of
the extended nature of the source on the image and the measured flux.
Such simulations have been conducted for ISGRI for different extended
source geometries, including \SO\ \citep{Renaud05sub}. For \SO,
\cite{Renaud05sub} used an input image of the expected \SY\ emission map in
20--40 keV band based on the simulations described in
\cite{Reynolds99}. The principle of the ISGRI simulations is the following:
For each point-like source constituting the extended one, the corresponding
shadowgram (the image of the mask pattern illuminated and projected onto
ISGRI) is calculated. The final expected shadowgram of \SO\ is obtained by
summing all these contributions. Then, the standard deconvolution
\citep[see][for details]{Goldwurm03} in OSA is applied to obtain the
reconstructed image. With this technique, \cite{Renaud05sub} obtained a
reduction factor of 0.7: that is, ISGRI would detect 70\% of
the true flux at each limb.



\epsscale{1.0}
\epsscale{1.0}

\section{Results}\label{sec:results}

\subsection{JEM-X results}\label{subsec:jemx}

The source is detected at the limbs in 2.4--4.2 keV and 4.2--8.4 keV
bands (see Fig.~\ref{fig:jemim} and
Table~\ref{table:jemxsum}). {\bf{This is the first time that the
structure of an individual extended source has been imaged with \integral.}}
It appears that the South-West (SW) limb (the limb on the right in the images)
is stronger in the 2.4--4.2 keV band at about the $1 \sigma$ level,
but the trend is reversed at higher energies. At 4.2--8.4 keV, the NE limb is
stronger, a result with higher significance.  An excess at the
position of the NE limb is present in the 8.4--14 keV band (1.7
$\sigma$), but no excess is seen in the SW limb. We note that \xmm\
data also points to an asymmetry in flux coming from the NE and SW
limbs such that the NE limb gets relatively stronger as energy
increases \citep{Rothenflug04}.  A similar trend is seen in \asca\
data \citep{Dyer04}, but the differences are small in the \asca\
band (below 8 keV).  The small excess of the SW over NE limbs in
the 2.4--4.2 keV band we see is not supported by \asca\
or \xmm\ observations in that energy range.

\begin{table}[t]
\caption{\label{table:jemxsum} JEM-X Summary}
\begin{minipage}{\linewidth}
\renewcommand{\thefootnote}{\thempfootnote}
\begin{tabular}{l|c|c|c} \hline \hline
{\bf{North-East Limb}} \\
Energy band & Flux & Statistical & Model flux\footnote
{Flux from extrapolated model fit to the \asca\ spectrum (0.8 - 9 keV), \citealt{Dyer04}} \\
(keV) & ($10^{-4}$ photon/cm$^2$s) & significance & ($10^{-4}$ photon/cm$^2$s) \\
\hline
2.4--4.2\footnote{The actual low energy threshold is variable over different parts of the JEM-X detector.} & 10 $\pm$ 4 & 2.6 & 14 \\
4.2--8.4 & 11 $\pm$ 2 & 5.0 & 5.9 \\
8.4--14 & 3.3 $\pm$ 1.9 & 1.7 & 1.4 \\
 & & &  \\
 {\bf{South-West Limb}} & & & \\
2.4--4.2 & 15 $\pm$ 4 & 4.0 & 13\\
4.2--8.4 & 3.6 $\pm$ 2.3 & 1.6 & 4.8 \\
8.4--14 & $<$5.7  & 3\footnote{3$\sigma$ upper limit} & - \\ \hline
\end{tabular}
\end{minipage}
\end{table}

\begin{figure}
\plotone{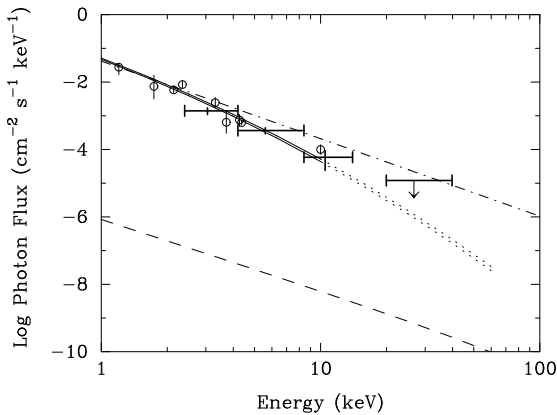} \caption{\label{fig:specpl} Integrated hard X-ray
spectrum of SN 1006. The heavy bars show the energy bins over which
our JEM-X fluxes are integrated. The integrated fluxes were
converted to spectral fluxes assuming a power-law index of 3.0, and
plotted at the median energy of each bin.  The errors in the bin
energies are insignificant (much smaller than the line widths). The
upper limit bar is from ISGRI. Open circles are pre-\asca\ (see
text) observations, and the two solid lines are the $\pm 1\sigma$
model fits to \asca\ and \rxte\ data, extrapolated to the higher
energies \citep{Dyer01}. The dotted-and-dashed line is the
extrapolation from the Chandra spectral fit \citep{Long03}. The
dashed line is the bremsstrahlung prediction of the model described
in text. }
\end{figure}

Figure~\ref{fig:specpl} shows our combined JEM-X fluxes for both
limbs, compared with previous spatially integrated pre-\asca\ flux
measurements (see \citealt{Reynolds96, Hamilton86} for references).
The integrated fluxes have been converted to flux densities assuming
a photon index $\Gamma$ of $3.0$ \citep{Allen01}. Also shown are a
pair of model curves for the escape model that provided a good fit
to \asca\ and \rxte\ data in \cite{Dyer04}. The parameters of the
best fit are a roll-off frequency of $\nu_{\rm roll} = 3.0 \times
10^{17}$ Hz and an electron energy index of $2.2$ (implying a radio
spectral index $\alpha = 0.6$ or radio photon index $\Gamma \equiv
\alpha + 1 = 1.6$). These parameters imply an e-folding energy of
the exponential cutoff (due to escape) of $32(B_2/10 \mu{\rm
Gauss})^{-1/2}$ TeV (where $B_2$ is the post-shock field). The
curves in Figure~\ref{fig:specpl} correspond to $\pm 1\sigma$ errors
on $\nu_{\rm roll}$: $(2.8 - 3.1) \times 10^{17}$ Hz. The dashed
line indicates the bremsstrahlung prediction for an upstream density
of 0.2 cm$^{-3}$ and $B_2 = 10 \mu$Gauss.

\subsection{ISGRI results}\label{subsec:isgri}

\begin{figure}
\plotone{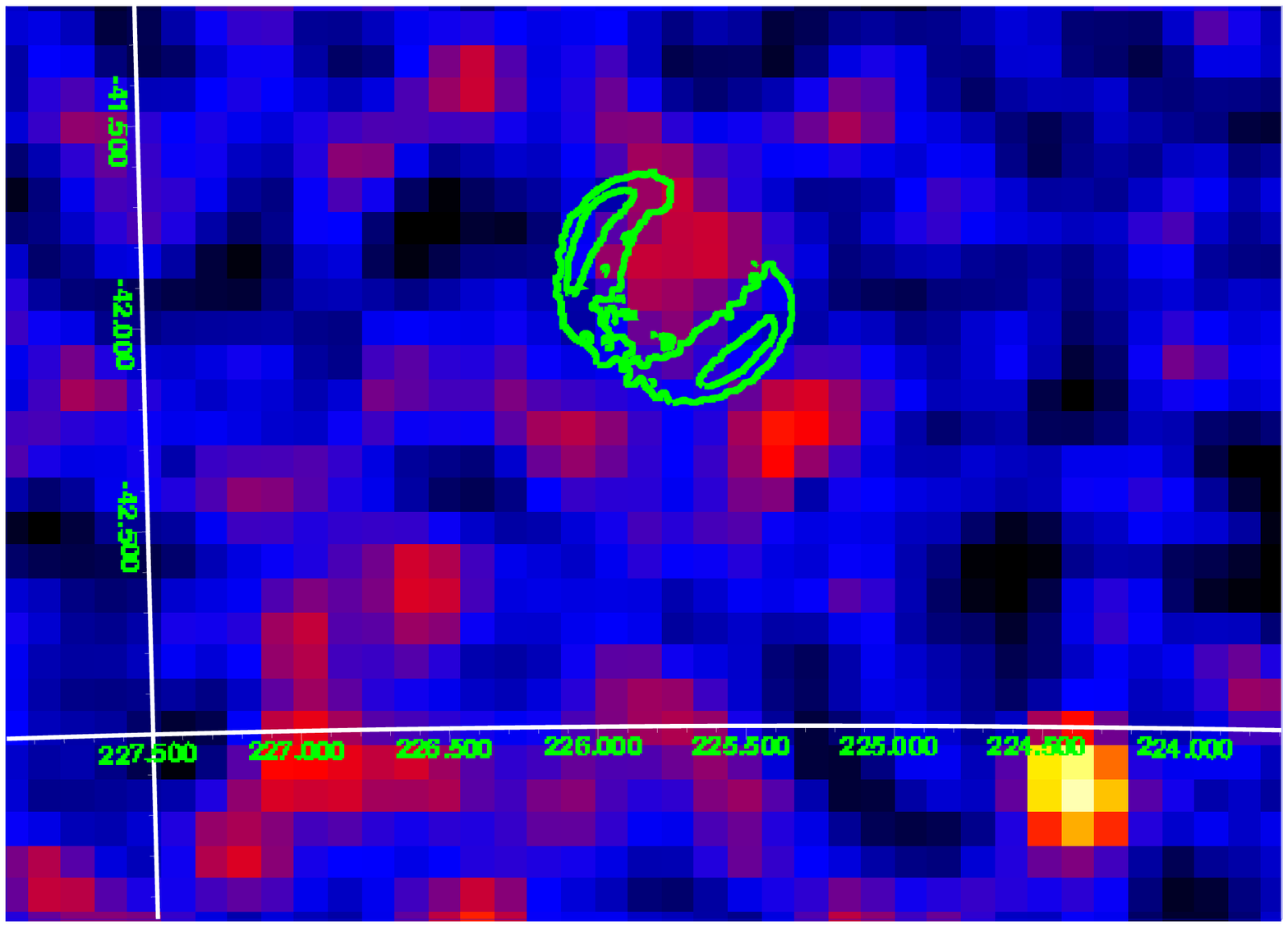} \caption{\label{fig:isgim} The ISGRI sigma image
around \SO\ in 20--40 keV band. \SO\ is not detected. A nearby
source \citep[possibly VV 780, 11.6 $\sigma$
detection][]{Kalemci05_atel} is also shown for comparison. The
\asca\ contours are overlaid to show the expected position of \SO.
The feature at the center of \SO\ is not significant, and is
possibly due to inaccurate background map. (The electronic version
has this figure in color.) }
\end{figure}

We obtained ISGRI images in different energy bands using OSA 4.2.
\SO\ was not detected in any band. The 20-40 keV sigma image is
shown in Fig.~\ref{fig:isgim}. The 3 $\sigma$ upper limit
(sensitivity limit) for a point source for \wsim 750 ks observing
time is \wsim$\rm 9 \times 10^{-5}$ photon cm$^{-2}$ s$^{-1}$. If
\SY\ dominates in 20--40 keV band, the emission would be
concentrated in two limbs. The extended nature of the source causes
a reduction factor of 0.7 in flux (see \S~\ref{subsec:isgrian}).
Therefore the 3 $\sigma$ upper limit for a synchrotron dominated
source is $\rm 1.3 \times 10^{-4}$ photon cm$^{-2}$ s$^{-1}$ at each
limb. For \SY\ dominated emission, the expected total flux in the
20--40 keV band (based on the models shown in
Figure~\ref{fig:specpl}) is $(2.9 - 3.6) \times 10^{-5}$ photon
cm$^{-2}$ s$^{-1}$ for the entire remnant, or roughly half this at
each limb. The bremsstrahlung emission from the model shown in
Figure~\ref{fig:specpl}, using the lower-limit 25 $\mu$G downstream
magnetic field from \emph{H.E.S.S.} TeV observations, is predicted
to be about $2 \times 10^{-8}$ photon cm$^{-2}$ s$^{-1}$, far below
ISGRI's sensitivity.  In fact, in this case, since the emission
would be coming from a larger area, the upper limit set by our
observations is even higher. The predicted crossover energy where
bremsstrahlung and synchrotron emission become comparable is in the
vicinity of 200 keV.

Long et al. (2003) reported that a power-law photon index of $2.30$
fit the Chandra data well between 0.5 and 5 keV for the emission at
the limbs. An extrapolation of this fit from 5 keV to the ISGRI
range is also shown in Fig.~\ref{fig:specpl} with a
dotted-and-dashed line. Our ISGRI upper limit is about a factor of 2
below the extrapolation of this power-law to 28 keV.

\section{Discussion}\label{sec:discussion}

Coded-mask imaging is a complex process, and optimal techniques are
still being developed to clean noisy images, both for JEM-X and ISGRI.  Our
results confirm and extend results obtained with \asca, \chandra, and \xmm,
and show that relatively small increases in sensitivity may allow ISGRI to
detect predicted synchrotron radiation for some models.

The prospect of detecting bremsstrahlung X-rays or gamma-rays from
SN 1006 has become considerably more remote with the combination of
lower estimates for the ambient density and the much higher magnetic
field that would be required to explain the lack of IC from CMB
photons in the \emph{H.E.S.S.} observations.  However, it is quite
possible that a somewhat longer \emph{INTEGRAL} observation
scheduled for Cycle 3 will have sufficient sensitivity to detect
synchrotron emission in the 18-40 keV band.  A model that can
describe the observations of Figure~\ref{fig:specpl} but with a
strong enough magnetic field to satisfy the \emph{H.E.S.S.}
constraints (downstream $B > 25 \mu$G) is somewhat harder than the
spectra shown in Figure~\ref{fig:specpl}, and predicts a flux in the
18-40 keV band of $0.3 \times 10^{-4}$ photon cm$^{-2}$ s$^{-1}$.
Some models predict considerably lower fluxes in the 18--40 keV
band, so ISGRI detection would not only extend and confirm the
presence of hard synchrotron X-rays, but could provide useful model
discrimination.

We have obtained the first observations with JEM-X of an extended
source, between 2.4 and 14 keV. The fluxes we derive are consistent
with those of earlier imaging observations where they overlap, and
support the identification of the continuum emission of SN 1006 as
synchrotron radiation from a slowly dropping off electron
distribution.  Our ISGRI 3$\sigma$ upper limit is about a factor of
4 higher than the prediction of the model that best fits the soft
X-ray continuum. With ISGRI, we confirm at higher energies than has
been previously reported that the X-ray spectrum of SN 1006 must be
steepening. A somewhat longer observation, and more developed data
analysis techniques, should allow the detection of SN 1006 in the
18-40 keV band with ISGRI, and above 8.4 keV with JEM-X, and can
provide important modeling constraints.


\acknowledgments E.K. is supported by the European Comission through
a FP6 Marie-Curie International Reintegration Grant (INDAM). E.K.
acknowledges partial support of T\"UB\.ITAK. E.K. and S.E.B.
acknowledge NASA grants NAG5-13142 and NAG5-13093. S.P.R.
acknowledges support from NASA grant NAG5-13092.



\clearpage






\end{document}